\documentclass [prl, twocolumn, aps, superscriptaddress
] {revtex4}
\usepackage{graphicx}
\usepackage{amsmath}
\usepackage{amssymb}

\newcommand{\vect}[1]{{\mathbf #1}}
\DeclareMathOperator{\tr}{Tr}

\begin{document}

\title{Quasi-one-dimensional polarized Fermi superfluids}

\author{Meera M. Parish}
\email{mparish@princeton.edu} \affiliation{Department of Physics,
Princeton University, Princeton,
NJ 08544} %
\affiliation{Princeton Center for Theoretical Physics, Princeton
University, Princeton, NJ 08544}

\author{Stefan K. Baur}
\affiliation{Laboratory for Atomic and Solid State Physics, Cornell
University, Ithaca NY 14853}

\author{Erich J. Mueller}
\affiliation{Laboratory for Atomic and Solid State Physics, Cornell
University, Ithaca NY 14853}

\author{David A. Huse}
\affiliation{Department of Physics, Princeton University, Princeton,
NJ 08544}

\date{\today}

\begin{abstract}
We calculate the zero temperature ($T=0$) phase diagram of a
polarized two-component Fermi gas in an array of weakly-coupled
parallel one-dimensional (1D) ``tubes'' produced by a
two-dimensional optical lattice. Increasing the lattice strength
drives a crossover from three-dimensional (3D) to 1D behavior,
stabilizing the Fulde-Ferrell-Larkin-Ovchinnikov (FFLO) modulated
superfluid phase. We argue that the most promising regime for
observing the FFLO phase is in the quasi-1D regime, where the atomic
motion is largely 1D but there is weak tunneling in the other
directions that stabilizes long range order. In the FFLO phase, we
describe a phase transition where the quasiparticle spectrum changes
from gapless near the 3D regime to gapped in quasi-1D.
\end{abstract}




\maketitle

Recent experiments on ultracold $^6$Li have probed polarized,
two-component Fermi gases as a function of interatomic interaction
strength and spin population
imbalance~\cite{3Dexpt}. 
These experiments have focussed on the unitarity regime, where the
$s$-wave scattering length is large and the scattering properties
are universal. At low temperatures, they have seen phase separation
between an unpolarized or weakly polarized superfluid phase and a
highly polarized normal phase.
Future experiments hope to observe the elusive FFLO
spatially-modulated superfluid phase, first predicted to occur in
magnetized superconductors over 40 years
ago~\cite{Fulde1964,larkin1965}, and having realizations in other
systems ranging from heavy-fermion superconductors~\cite{bianchi03}
to quark matter~\cite{alford01}.
While the FFLO phase is expected to exist in trapped 3D gases for
small polarizations and weak attractive
interactions~\cite{Mizushima2005},
it is predicted to only occupy a small part of the $T=0$ phase
diagram~\cite{sheehy2006}, and this region is only further
diminished with increasing temperature~\cite{FFLO_T}.
Here we theoretically explore how a two-dimensional (2D) optical
lattice can enlarge the region of stability of the FFLO phase,
paving the way for its observation.

Although a 3D simple cubic optical lattice may also enhance the FFLO
phase~\cite{koponen2007}, we argue that a 2D optical lattice, which
permits free motion in one direction, is more effective. Analogous
to charge density wave instabilities, the instability of the normal
state to FFLO is due to a Fermi surface ``nesting'', which is
enhanced in 1D. By increasing the intensity of the optical lattice,
one can continuously tune the single atom dispersion from 3D to 1D,
a scenario which is readily achieved experimentally~\cite{moritz05}.

As revealed by the Bethe \textit{ansatz}~\cite{orso2007,hu2007}, the
exact $T=0$ phase diagram of the 1D polarized Fermi gas displays
four phases: unpolarized superfluid (SF), FFLO, fully-polarized
normal, and vacuum, characterized by the densities of the two
species and by algebraic order. Unlike 3D, in 1D all of the phase
transitions are continuous and the FFLO phase occurs at \textit{all}
non-zero partial polarizations for \textit{any} strength of the
attractive interaction~\cite{hu2007}. Furthermore, at the phase
boundary between polarized and unpolarized phases, the SF phase has
the lower density in 1D, which is opposite to the situation in 3D.
In a trap, the spatial structure is therefore inverted: in 1D one
has a central FFLO region surrounded by SF, for small polarizations.
During the crossover from 1D to 3D, we find a regime where the phase
sequence moving from the center to the edge of the trap is the
quasi-1D FFLO, then a shell of SF, then the more 3D-like polarized
normal.

We consider a gas of fermionic atoms in two different hyperfine
states (labeled by $\sigma = \uparrow,\downarrow = \pm1$) confined
by a smooth trapping potential $V({\bf r})$ and a square optical
lattice potential such as $V_L({\bf r})=-U ({\bf r}) \left[\cos(2\pi
x/b)+\cos(2\pi y/b) \right],$ with lattice spacing $b$ and local
depth $U({\bf r})$, which breaks the cloud into an array of tubes
aligned along the $z$-direction. In the local density approximation
(LDA), the properties at location ${\bf r}$ depend on $U({\bf r})$
and the local chemical potentials $\mu_\sigma(\vect{r}) \equiv
\mu_\sigma - V(\vect{r})$ in the same way that a spatially uniform
system does. Within LDA, which should be valid for a wide enough
trap, the spatially varying pattern of phases in a trap can be read
off from the phase diagram of the uniform system by tracing the
spatial variation of $\mu \equiv
(\mu_{\uparrow}+\mu_{\downarrow})/2$ and $U$, while holding the
difference $h \equiv (\mu_{\uparrow}-\mu_{\downarrow})/2$ fixed. In
the special case of uniform $U$, only $\mu$ varies in space.

To produce the uniform phase diagram we study the untrapped system
in a uniform lattice with $N_x \times N_y$ tubes, each of length
$L_z$ in the $z$-direction. For a sufficiently low density and
strong enough lattice, the $xy$ motion is well approximated by a
one-band tight-binding model with single-atom dispersion:
\begin{equation} \label{eq:disp}
 \epsilon_{\vect{k}} = \frac{k_z^2}{2m} + 2t[2 - \cos(k_xb) -
 \cos(k_yb)]~,
\end{equation}
where $t$ is the hopping (related to $U$ and $b$ as in
\cite{jaksch}), $m$ is the atomic
mass, and we use $\hbar = 1$. The 
Brillouin zone of the $xy$ motion is $|k_x|, |k_y|\leq\pi/b$, while
$k_z$ is unconstrained. For energies well above the $xy$ bandwidth
 $\epsilon_{\vect{k}} \gg 8t$, the dispersion is 1D-like.
For low energies $\epsilon_{\vect{k}} \ll t$, the dispersion is
3D-like, and can be made isotropic if we rescale the momenta $\{k_x,
k_y, k_z\} \mapsto \{b\sqrt{2t}k_x, b\sqrt{2t}k_y, k_z/\sqrt{m}\}$.
This single-band, tight-binding regime is accessed experimentally by
working in a regime with $t, \varepsilon_{F\sigma} \ll
\sqrt{U/mb^2}$, where $\varepsilon_{F\sigma} = (\pi n_\sigma
b^2)^{2}/2m$ is the 1D Fermi energy for each species of density
$n_\sigma$ (corresponding to the 1D density $n_\sigma b^2$ per tube
in the optical lattice).

Since the $^6$Li experiments use highly dilute gases with a wide
Feshbach resonance, the interactions can be modeled by a contact
interaction, giving a Hamiltonian,
\begin{multline} \label{eq:ham}
  \hat{H} - \mu_{\uparrow} \hat{N}_{\uparrow} - \mu_{\downarrow} \hat{N}_{\downarrow}
  = \sum_{\vect{k}} \sum_{\sigma = \uparrow,\downarrow}
  \left(\epsilon_{\vect{k}} - \mu_{\sigma}\right)
  c_{\vect{k} \sigma}^\dag c_{\vect{k} \sigma} \\
  + \frac{g}{L_z N_x N_y}
  \sum_{\vect{k},\vect{k}',\vect{q}}
  c_{\vect{k} \uparrow}^\dag c_{\vect{k}'
  \downarrow}^\dag c_{\vect{k}'+\vect{q} \downarrow}
  c_{\vect{k}-\vect{q} \uparrow} ~.
\end{multline}
%
%
Solving the 3D scattering problem in a single harmonic tube of
transverse size $a_{\perp} = \sqrt{1/m\omega_{\perp}}$, one
finds~\cite{olshanii1998}:
\begin{equation}\label{eq:int}
  \frac{1}{g} = - \frac{ma_{1D}}{2} \equiv \frac{ma_{\perp}}{2}
  \left(\frac{a_{\perp}}{a} - C \right)
\end{equation}
where $a$ is the 3D $s$-wave scattering length, arising from the
short-range interatomic potential.
Thus, we have an attractive 1D interaction when $a_{\perp}/a < C
\simeq 1.4603/\sqrt{2}$. Defining the 1D two-body binding energy
$\varepsilon_B = g^2m/4$, we can fully parameterize the $T = 0$
phase diagram of the uniform system with three dimensionless
quantities: $t/\varepsilon_B$, $\mu/\varepsilon_B$ and
$h/\varepsilon_B$~\cite{param}.

We calculate the $T=0$ phase diagram within mean-field theory, which
captures most of the qualitative features of the phase diagram as we
move between the 1D and 3D regimes. However, we know by comparison
to the exact solution that this mean-field approximation does miss
some features of the 1D limit, as we note below.

\begin{figure}
\centering
\includegraphics[width=0.4\textwidth]{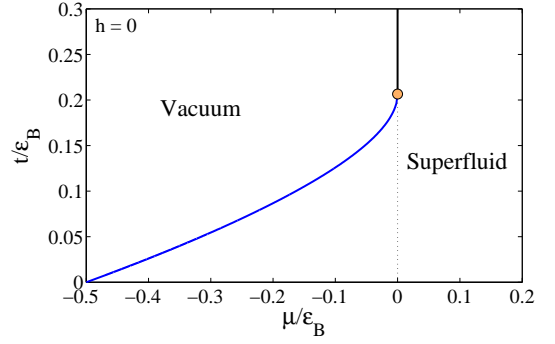}
\caption{(Color online) Phase diagram for $h=0$. For
$t/\varepsilon_B$ below the filled circle, there is a two-atom bound
state, and the resulting bosonic pairs enter the system as a BEC as
$\mu$ is increased through the solid line (given by $\mu = -E_B/2$).
For $t/\varepsilon_B$ above the filled circle we are always in the
BCS regime.}\label{fig:h0}
\end{figure}

We begin at $h = 0$ where there are only two phases: SF and the
vacuum (see Fig.~\ref{fig:h0}). In the 1D limit (small $t$) there is
a two-atom bound state with binding energy $E_B$, where clearly $E_B
= \varepsilon_B$ when $t = 0$. These bosonic pairs enter the system
and form a Bose-Einstein condensate (BEC) for $\mu
> -E_B/2$. Increasing the density further brings the system through
a density-driven BEC-BCS crossover, similar to excitonic
systems~\cite{Comte}, where $\mu/E_B \gg 1$ defines the
weak-coupling BCS limit.
Making the system more 3D, by increasing $t$, reduces both $E_B$ and
the BEC regime. For $t/\varepsilon_B > 0.2066$ there is no two-atom
bound state and thus only the BCS regime.

\begin{figure}
\centering
\includegraphics[width = 0.48\textwidth]{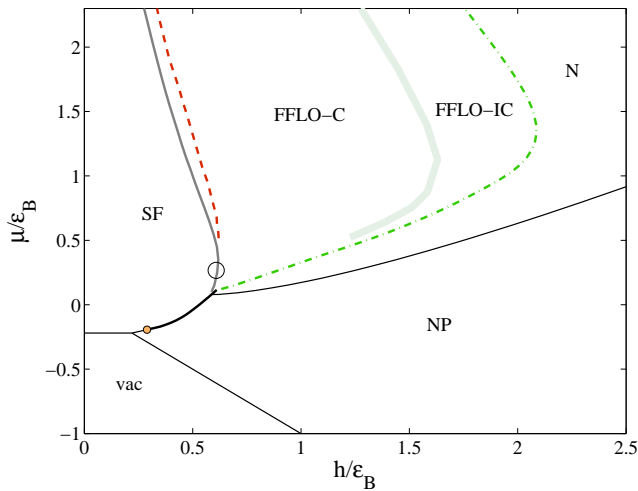}
\caption{(Color online) Slice of the mean-field phase diagram taken
at $t/\varepsilon_B = 0.08$. The phases shown include the
unpolarized superfluid (SF), partially-polarized normal (N), and
fully-polarized normal (NP). The FFLO phase is divided into gapped
`commensurate' (C) and ungapped `incommensurate' (IC) phases. The
filled circle marks the tricritical point; near it, but not visible
here is a tiny region of SF$_{\textrm{M}}$ magnetized superfluid
phase, a remnant of the 3D BEC regime.  The SF-NP and SF-N
transitions are first-order for $\mu/\varepsilon_B$ above the
tricritical point, along the solid heavy line. The SF-FFLO
transition (solid line) is estimated from the domain wall
calculation. The transition from FFLO to normal (dotted-dashed line)
is assumed to be second-order. The large circle marks the region of
FFLO where $\Delta/\varepsilon_F$ is largest, so the phase is likely
to be most robust to $T>0$ here. The dashed line near the SF-FFLO
transition shows where the wave vector of the FFLO state is
stationary as a function of $\mu$: $dq/d\mu=0$ (this is calculated
using the FF approximation).}\label{fig:tslice}
\end{figure}

At finite $h$ we must also consider the FFLO superfluid phase, where
the Cooper pairs condense with nonuniform pairing order parameter
(gap)
\begin{eqnarray} \notag 
 \Delta(z)  & = &
 - \frac{g}{L_z N_x N_y}
  \sum_{{\bf k},q_z} e^{i q_z z}
\langle c_{-{\bf k}+q_z {\bf \hat z}/2\downarrow}c_{{\bf k}+q_z {\bf
\hat z}/2\uparrow}\rangle
\\
& = &
 -\frac{g}{N_x N_y}
 \sum_{l,\mathbf{k_{\perp}}}  u_{l,\mathbf{k_{\perp}}}(z)
 v^*_{l,\mathbf{k_{\perp}}}(z) f( E_{l,\mathbf{k_{\perp}}}),
\end{eqnarray}
where $f(x)$ is the zero-temperature Fermi function,
$\vect{k_{\perp}}$ is the momentum in the $xy$ plane, $l$ labels the
quasiparticle modes, and the energies/coherence factors obey the
Bogoliubov-de Gennes (BdG) equations \cite{degennes}
\begin{eqnarray}
\label{bdg_equations} \left(
\begin{array}{cc}
h_{0\uparrow} 
& \Delta(z) \\
\Delta^*(z) & -h_{0\downarrow} 
\end{array}
\right) \left(
\begin{array}{c}
u_{l,\mathbf{k_{\perp}}}(z) \\
v_{l,\mathbf{k_{\perp}}}(z)
\end{array}
\right)  = E_{l,\mathbf{k_{\perp}}} \left(
\begin{array}{c}
u_{l,\mathbf{k_{\perp}}}(z) \\
v_{l,\mathbf{k_{\perp}}}(z)
\end{array}
\right)
\end{eqnarray}
with $h_{0\sigma} = -\frac{\partial_z^2}{2 m}+ 2t[2 - \cos(k_xb) -
\cos(k_yb)] + g n_{-\sigma}(z)-\mu$. The densities in the Hartree
term are $n_\uparrow(z)=(1/N_x N_y)\sum_{l,\mathbf{k_{\perp}}}
|u_{l,\mathbf{k_{\perp}}}(z)|^2 f(E_{l,\mathbf{k_{\perp}}})$ and
$n_\downarrow(z)=(1/N_x N_y)\sum_{l,\mathbf{k_{\perp}}}
|v_{l,\mathbf{k_{\perp}}}(z)|^2 f(-E_{l,\mathbf{k_{\perp}}})$. The
grand potential is ~\cite{PhysRev.187.556}
\begin{eqnarray}
\label{freeenergy}
&&\Omega =
- N_x N_y \int dz \left[\frac{|\Delta(z)|^2}{g} + g n_{\uparrow}(z)
n_{\downarrow}(z)\right]\\\nonumber&&\,\,+
\tr\left[\frac{h_{0\uparrow}+h_{0\downarrow}}{2}
\right] +\!\sum_{l,\mathbf{k_{\perp}},\sigma}
\frac{E_{l,\mathbf{k_{\perp}}}+\sigma h}{2}
f(E_{l,\mathbf{k_{\perp}}}+\sigma h).
\end{eqnarray}
where the sum includes both positive and negative energy
eigenvalues.
The simplest {\it ansatz} for the FFLO phase is Fulde and Ferrell's
one-plane wave state $\Delta(z) = \Delta_{FF} e^{i q
z}$~\cite{Fulde1964}. Here, the energy eigenvalues
$E_{l,\mathbf{k_{\perp}}}$ reduce to $E_{\vect{k}\pm} = \pm \sqrt{
(\xi_+^\uparrow+\xi_-^\downarrow)^2/4 + \Delta^2} +
(\xi_+^\uparrow-\xi_-^\downarrow)/2 $, with $\xi^\sigma_{\pm} =
\epsilon_{\vect{k} \pm q \hat{\vect{z}}/2}-\mu+g n_{-\sigma}$ and we
can then minimize Eq.~(\ref{freeenergy}) directly. This state is a
good approximation in the limit $\Delta_{FF} \rightarrow 0$. Indeed,
one can show that the second-order transition to the normal phase
occurs at single wave vector $q$ when~\cite{sec_order}
\begin{eqnarray}\label{eq:thouless}
-\frac{1}{g}&=&\frac{1}{N_x N_y L_z}\sum_{\bf k}
\frac{1-f(\xi^\uparrow_+-h)-f(\xi^\downarrow_-+h)}{\xi^\uparrow_+
+\xi^\downarrow_-}.
\end{eqnarray}
The locus of this transition is illustrated in
Fig.~\ref{fig:tslice}.

Larkin and Ovchinnikov \cite{larkin1965} showed that the energy is
lower if the Cooper pairs condense in a standing wave, with
$\Delta(z)=\Delta_{LO} \cos(q z)$ when the gap is small. More
generally, $\Delta(z)$ is a real periodic function of $z$.
When the coherence length $\xi$ is small compared to $1/q$, this
state consists of well-separated domain walls between domains where
$\Delta$ is alternately positive and negative. The polarized cores
of these domain walls result from occupying the spin-up Andreev
bound states on each wall~\cite{yoshida:063601}.

We calculate the energy of a single domain wall by iterating to
self-consistency Eq.~(\ref{bdg_equations}) in a finite box with
periodic boundary conditions, beginning with a trial $\Delta(z)$
containing two domain walls whose separation is large compared to
the coherence length. If the domain walls interact repulsively, the
SF to FFLO transition is continuous and lies where this domain wall
energy vanishes; otherwise this condition marks the spinodal of a
first-order transition (likely to be near the true phase boundary).
Within mean field theory the transition is continuous in 1D
\cite{PhysRevB.30.122}, and has been argued to be so in 3D
\cite{matsuo,yoshida:063601,yang2001}: in weak coupling the critical
fields are respectively $h=(2/\pi)\Delta_0(=0.64 \Delta_0),$ and
$0.67 \Delta_0$, where $\Delta_0$ is the gap in the SF phase. We are
unaware of a strong coupling 3D calculation of the sign of the
domain wall interaction.

Fig.~\ref{fig:tslice} shows a representative slice of the mean-field
phase diagram at fixed $t/\varepsilon_B=0.08$ (if one can neglect
the spatial variation of $\epsilon_B$ and $t$, this slice
corresponds to a fixed optical lattice intensity). Near the vacuum
at small filling (low $\mu$) is the 3D BEC regime, including a very
small region of the SF$_M$ magnetized superfluid phase where the
excess fermions form a Fermi liquid within the BEC. As $\mu$ and
thus the filling is increased, the system crosses over towards 1D.
Here, the FFLO phase appears and occupies a large portion of the
phase diagram~\cite{3D_FFLO}. Both the SF and FFLO phases become
re-entrant: in the 1D regime the FFLO phase is at a higher $\mu$ and
thus a higher density than SF, while in the 3D regime this density
relation is reversed.
Thus, we see that the ``inverted'' phase separation in 1D trapped
gases is connected to the standard phase separation of 3D via an
intermediate pattern of phases where SF forms a shell surrounded by
polarized phases.
As $t/\varepsilon_B$ is further reduced, the 3D regime becomes
smaller, with the re-entrance of the SF phase moving to lower $\mu$,
while the FFLO phase grows and the sliver of N phase between FFLO
and NP is diminished.  In the limit $t=0$ this phase diagram matches
fairly well to that obtained from the exact solution in 1D (e.g.,
Fig. 1 of Ref.~\cite{orso2007}).  The main feature that the
mean-field approximation misses at $t=0$ is the multicritical point
where the four phases, SF, FFLO, NP and vacuum, all meet at
$h=-\mu=\varepsilon_B/2$.  In mean-field theory, the FFLO phase
never extends all the way down to zero density; instead it is
preempted by a first-order SF-to-NP transition.

A new $T=0$ phase transition occurs within the FFLO phase as one
moves from 3D to 1D by increasing the intensity of the 2D optical
lattice.  In 3D the FFLO state has a Fermi surface, and is therefore
gapless.  In 1D the spectrum of BdG quasiparticles is fully gapped
in the FFLO state. The gapped, commensurate FFLO state (FFLO-C)
contains exactly one excess spin-up atom per 1D tube per domain
wall.  This commensurability means that $q=\pi
b^2(n_{\uparrow}-n_{\downarrow})$, while, by contrast, the number of
excess up spins in the ungapped, incommensurate FFLO state (FFLO-IC)
is not constrained.

The transition between FFLO-C and FFLO-IC can be understood from the
band structure of the Andreev bound states on the domain walls.  In
FFLO-C the chemical potential lies in a gap in the quasiparticle
spectrum. Thus, the superfluid FFLO-C phase is a band insulator for
the {\it relative} motion of the unpaired atoms and the condensate
of pairs. As the optical lattice intensity is decreased, the 3D
bands broaden and may overlap the chemical potential, opening up a
Fermi surface. We approximate the IC-C transition within the FF {\it
ansatz} by examining the $k_z > 0$ half of the Fermi surface to see
if it is fully gapped. In the limit $\mu/t \gg 1$, the transition
occurs when $\Delta \sim 8th/\mu$.

We now address the question of what are the best conditions for
experimentally producing, detecting and studying the FFLO phase.
Ideally one might use {\it in situ} imaging to directly observe the
spatial density and magnetization modulations in this phase. In a
trapped 3D gas, the modulated superfluid will occupy a hard to
detect thin shell. The thinness of this shell results from the small
range of $\mu$ over which the FFLO phase is
stable~\cite{sheehy2006}. Even approaches which produce an enlarged
FFLO region in density space~\cite{yoshida:063601} 
share this feature. Moreover, imaging the modulations will be
complicated by their 3D nature (e.g., they may form an onion-like
pattern). The 1D limit also has problems. Although in 1D the FFLO
phase occurs in a large region of the $T=0$ phase diagram, it has no
true long-range order, only power-law correlations. Furthermore, the
transition temperature ($T_c$) of this 1D superfluid is zero. Also,
for $t=0$ (strictly 1D) we have an array of independent parallel 1D
clouds whose density modulations will be out of phase with one
another, reducing the observed signal.

Given these concerns, we believe that the optimal conditions for
observing FFLO are likely to be in the quasi-1D regime, where the 2D
optical lattice is at an intermediate intensity that is strong
enough to make the Fermi surface 1D-like (and hence enhance the
instability towards FFLO), but weak enough that the atoms are still
able to hop between the parallel tubes and thus introduce strong
inter-tube correlations in the optical lattice. The resulting 3D
long-range order can then survive to nonzero
temperature~\cite{yang2001}.

Although we have only performed a $T=0$ calculation, we can crudely
estimate $T_c$ from the size of the gap $\Delta$. For small
$\Delta$, superfluid phases have $T_c\propto \Delta$, but when
$\Delta$ approaches $\varepsilon_F$, $T_c$ saturates. Thus the
observability of a superfluid phase such as FFLO is enhanced if the
gap is increased to of order $\varepsilon_F$, but there is not
likely to be an advantage to increasing the gap to much larger
values. In 3D the maximum $T_c/\varepsilon_F$ of the SF occurs on
the BEC side of the Feshbach resonance, well away from the FFLO
phase \cite{sademelo1993}. However, as we move towards 1D in our
phase diagram, the FFLO phase extends more and more into the regime
of strong pairing where the gap is of order $\varepsilon_F$, and
thus we expect a large $T_c$. For a given $t/\varepsilon_B$, we find
that the gap in the FFLO phase is the largest fraction of
$\varepsilon_F$ at the SF-FFLO phase boundary near its point of
re-entrance, where $h$ on the SF-FFLO boundary reaches its maximum
(see Fig. 2).  We also find that within mean-field theory this
fraction $\Delta/\varepsilon_F$ increases as we reduce the hopping
$t$.  At sufficiently low $t$ the system crosses over from quasi-1D
to 1D and our mean-field theory becomes unreliable.  In the 1D
limit, $T_c$ must vanish, so the maximum value of
$T_c/\varepsilon_F$ within the FFLO phase must occur in the quasi-1D
regime at some small but nonzero hopping $t$.

Another consideration that may complicate the detection of the FFLO
phase within a trap is the fact that $\mu$ varies spatially both
within and between tubes. This means that the local wavenumber $q$
of the modulation will vary through the cloud, making the
modulations more difficult to detect. However, this variation can be
minimized if one works near a point where $dq/d\mu=0$.  We find such
points do exist in the quasi-1D regime (see Fig.~\ref{fig:tslice});
in 3D, $dq/d\mu$ is always negative so such points do not exist.
Note that in the 1D limit ($t=0$) there is even a point in the exact
phase diagram near strong coupling where $d^2q/d\mu=dq/d\mu=0$ that
should be a real ``sweet spot'' for having a uniform $q$ over a
fairly large fraction of a trap, and that this feature should
survive to small $t$.

We thank Randy Hulet for many helpful discussions.  This 
work was supported in part by the Army Research Office Grant No.
W911NF-07-1-0464.


\begin{thebibliography}{25}
\bibitem{param}
Only 2 parameters are needed in 3D, e.g. $\mu/h$, $1/a\sqrt{2mh}$.
\end{thebibliography}

\end{document}